# Ultra-high density alignment of carbon nanotubes array by dielectrophoresis

Shashank Shekhar[1,2], Paul Stokes[1,2], and Saiful I. Khondaker [1,2,3]*

[1] Nanoscience Technology Center, [2] Department of Physics, [3] School of Electrical Engineering and Computer Science, University of Central Florida, Orlando, Florida 32826, USA.

Abstract

We report ultra-high density assembly of aligned single walled carbon nanotubes (SWNTs) two dimensional arrays via ac dielectrophoresis using high quality surfactant free and stable SWNT solutions. After optimization of frequency and trapping time, we can reproducibly control the linear density of the SWNT between prefabricated electrodes from 0.5 SWNT/µm to more than 30 SWNT /µm by tuning the concentration of the nanotubes in the solution. Our maximum density of 30 SWNT/µm is the highest for aligned arrays via any solution processing technique reported so far. Further increase of SWNT concentration results dense array with multiple layers. We discuss how the orientation and density of the nanotubes vary with concentrations and channel lengths. Electrical measurement data show that the densely packed aligned arrays have low sheet resistances. Selective removal of metallic SWNTs via controlled electrical breakdown produced field effect transistors (FET) with high current on-off ratio. Ultra-high density alignment reported here will have important implications in fabricating high quality devices for digital and analog electronics.

* To whom correspondence should be addressed. E-mail: saiful@mail.ucf.edu

1. **Introduction:**

Single-walled carbon nanotubes (SWNTs) are considered to be a promising building block for future digital and analog electronic circuits due to their exceptional electronic properties [1,2]. Electron transport measurements of devices fabricated from individual SWNTs have displayed sub-threshold swings as low as 60 mV/decade, mobilities reaching 79000 cm$^2$/V s and conductance nearing the ballistic limit ($G = 4e^2/h$ ~155 μS or $R$ ~6.5 kΩ) [3,4]. One of the main challenges in nanotube electronics is the chirality control of the individual SWNT which causes large device to device inhomogenity in performance. Devices fabricated from arrays of SWNTs can be advantageous over individual tube devices, as they may provide more homogenity from device to device and can cover large areas. In addition, devices fabricated with nanotube arrays contain hundreds of parallel SWNTs contributing in charge transport, which can increase current outputs up to hundreds of microamperes. Aligned arrays of SWNT are also of particular interest in radio frequency (RF) applications as it may provide the higher cut-off frequency in THz regime and low input impedance (~ 50 Ω) [5]. In addition, the degree of alignment and the density of SWNTs have a great influence on device performance fabricated with these aligned array nanotube. Perfectly aligned array of SWNTs is expected to exhibit the electronic property that approaches the intrinsic properties of individual nanotube. Due to these advantages, there is a push to fabricate devices with massively parallel arrays of SWNTs [5-25]. Such devices may be useful for radiofrequency applications [5-7,15], transistors [10], plastic electronics [10-12], display technologies[27-29] and sensors [30].

Several techniques are being developed for the fabrication of dense aligned arrays of SWNT. Direct growth via chemical vapor deposition (CVD) using pattern lines of catalyst on single crystal quartz or sapphire substrate achieved aligned array with an average linear density of ~10 SWNT/μm using single growth technique [10,17] and ~30 SWNT/μm using more complicated double growth technique [18]. However, direct growth technique requires very high temperature (~900 $^0$C) and transfer printing is required to transfer the aligned array to suitable substrates for device fabrication [6,10,14]. Post growth assembly using solution processed SWNT could be advantageous due to its ease of processing at room temperature, compatibility with current Complementary Metal Oxide Semiconductor fabrication, and potential for scaled up manufacturing of SWNT devices on various substrates. Among the post growth techniques, langmuir-blodgett assembly [19], bubble blown assembly [20], evaporation driven self assembly [22], spin coating assisted alignment [23] and contact printing [24] have demonstrated reasonably high density alignment with ~ 1 SWNT/μm in most cases and up to 10 SWNT/μm in few cases.

One technique that has the potential to assemble SWNT with ultra-high density alignment is AC dielectrophoresis (DEP) [8, 25, 31]. DEP has been used to assemble 2D, 1D and 0D nanomaterials at the selected position of the circuit for device applications [32-36]. DEP can be advantageous over other solution processed techniques because it allows the materials to be directly integrated to prefabricated electrodes at the selected positions of the circuits and does not require post etching or transfer printing. One crucial aspect of the DEP process for high density alignment is the quality of the SWNT solution. The solution should be free of catalytic particles, contain mostly individual SWNTs, and be stable for long periods of time. Catalytic particles in the solution tend to make their way into the electrode gap with the SWNTs during assembly process due to their highly conductive nature which can disrupt the assembly and degrade device performance. Solutions containing bundles makes it difficult to obtain only individual SWNTs reproducibly into the electrode gap as the DEP force will likely select the larger bundles due to their higher dielectric constant and conductivity.

In this paper, we used a clean, surfactant free and stable SWNT aqueous solution combined with the DEP technique to achieve ultra-high density alignment of SWNTs in an array. The assembly was done by applying an AC voltage of 5 Vp-p at 300 KHz between prefabricated palladium (Pd) source and drain electrodes of channel lengths (L) of 2, 5 and 10 µm while channel width (W) was varied from 25 µm to 1 mm to show the scalability of the process. By simply tuning the concentration of the nanotubes in the solution, we can reproducibly control the linear density of the SWNT uniformly in the two dimensional array from 0.5 SWNT/µm to more than 30 SWNT/µm. Our maximum density of 30 SWNT/µm is the highest reported density of aligned arrays via any solution processing technique. Further increase of SWNT concentration results dense array with multiple layers. We discuss how the orientation and density of the nanotube vary with concentrations and channel lengths. Electrical measurement data show that the densely packed aligned arrays have sheet resistances of 3-10 K$\Omega\square^{-1}$. Selective removal of metallic SWNTs via controlled electrical breakdown produced field effect transistors (FET) with high current on-off ratio. Ultra-high density alignment using solution processed SWNT reported here will have important implication in fabricating high quality nanoelectronic devices.

## 2. Experimental Details

Figure 1a shows a cartoon of the DEP assembly of setup. The channel lengths (L) were 2, 5, and 10 µm while the channel width (W) was varied from 25 µm to 1 mm. The directed assembly of SWNTs at predefined electrodes was done in a probe station under ambient conditions via DEP. The SWNT aqueous solution was purchased from commercial source, was free from surfactant and catalytic particles and was stable for at least six months [37]. The solution has an original SWNT concentration of ~ 50 µg/ml and was diluted using de-ionized (DI) water to a desired concentration. The average diameter of the nanotubes were 2 nm and the length of the nanotubes varied from 0.5 to 10 µm with a median value of 1-2 µm as determined from atomic force microscopy (AFM) and scanning electron microscopy (SEM) investigations. Prior to the assembly, the electrodes were placed in oxygen plasma cleaner for 10 minutes to remove any unwanted organic residues on the surface. For the assembly, a small (3 µL) drop of the SWNT solution was cast onto the chip containing the electrode arrays. An AC voltage of 5 Vp-p with frequency 300 KHz, was applied using a function generator between the source and drain electrodes for 30 seconds. The AC voltage gives rise to a time averaged dielectrophoretic force. For an elongated object it is given by $F_{DEP} \propto \varepsilon_m \text{Re}[K_f] \nabla E_{RMS}^2$,

$$K_f = \frac{\varepsilon_p^* - \varepsilon_m^*}{\varepsilon_m^*}, \quad \varepsilon_{p,m}^* = \varepsilon_{p,m} - i\frac{\sigma_{p,m}}{\omega},$$ where $\varepsilon_p$ and $\varepsilon_m$ are the permittivity of the nanotube and solvent respectively, $K_f$ is the Claussius-Mossotti factor, $\sigma$ is the conductivity, and $\omega = 2\pi f$ is the frequency of the applied AC voltage [38]. The induced dipole moment of the nanotube interacting with the strong electric field causes the nanotubes to move in a translational motion along the electric field gradient and align in the direction of the electric field lines. The DEP assembly depends mainly on four parameters namely, applied external voltage (Vp-p), sinusoidal frequency (f), concentration of SWNT solution and DEP time. In principle the controlled alignment can be achieved by optimizing any of the above parameters. We found that the optimum nanotube assembly occurs between 300 kHz -1 MHz and for 5 – 10 V. We observed that waiting for a period of 30 seconds resulted in homogeneous assembly and a better reproducibility.

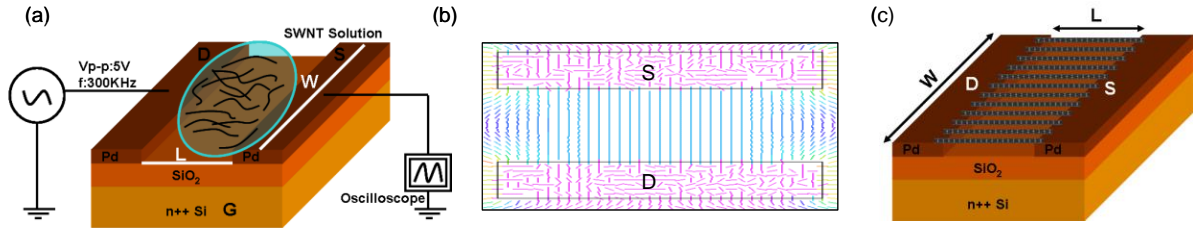

Figure 1. (a) Schematic of the DEP assembly. (b) 2D simulated electric field around the electrode gap. S and D are Pd source and drain electrodes. (c) Cartoon of the aligned nanotubes after DEP process. By varying the concentration of SWNT solution, the number density of the assembled nanotubes can be tuned.

Figure 1b shows a simulation of the electric field around the electrode gap for the parallel electrode patterns. The simulations were done using a commercially available software (Flex PDE) assuming that the potential phasor is real and therefore using the electrostatic form of the Laplace equation ($\nabla^2 \Phi = 0$). Hence we can set the effective potential of the electrodes to $\Phi = \pm V_{p-p}/2$ for our simulation. As can be seen from the simulation, the electric field is uniform throughout the electrode gap allowing for many nanotubes to align parallel to one another throughout the gap. After the assembly, the function generator was turned off and the sample was blown dry by a stream of nitrogen gas. Figure 1c shows schematic of an array of assembled nanotubes.

### 3. Results and Discussion:

Figure 2 shows SEM images for a typical DEP assembly when the concentration of the SWNT solution was varied by simply diluting the original solution using DI water. The electrode pairs have L= 2 µm and W= 25 µm. When the SWNT concentration was 0.08 µg/ml (diluted 600 times), on an average there was 1 SWNT/µm assembled between the electrodes as shown in figure 2a. Figure 2b, shows image where the concentration was 0.34 µg/ml (diluted 150 times). From this image, we count an average of 10 SWNT/µm entirely bridging the two electrodes. As the concentration of the solution is increased to 1.67 µg/ml (30 times dilution) and 3.4 µg/ml (15 times dilution), we found that the linear density of the SWNT is increased to ~20 and ~30 /µm as shown in figure 2c and 2d respectively. A magnified view of these images is shown in figure 2e and 2f. At low densities, almost all of the nanotubes are well aligned and are parallel to each other. However, at higher density, about 90% are aligned within $\pm 10^0$ of the longitudinal axis (figures 2e and 2f). As can be seen from these images, by simply varying the concentration of the SWNT solution, we are able to vary the density of the SWNT in the channel from 1 to ~30 SWNT/um. The maximum density reported here is the highest for any solution processed technique and is comparable to double growth CVD technique which was done at very high temperature (900 $^0$C) [18]. One crucial aspect of our alignment is that the density of the nanotube is uniform throughout the channel width. This uniformity of the SWNT can be attributed to the uniform electric field lines in a parallel plate arrangement. In addition, in contrast to previous DEP assemblies our aligned nanotube array does not contain any bundle nor does it contain any catalytic particle. This is due to the high quality SWNT solutions used in this study.

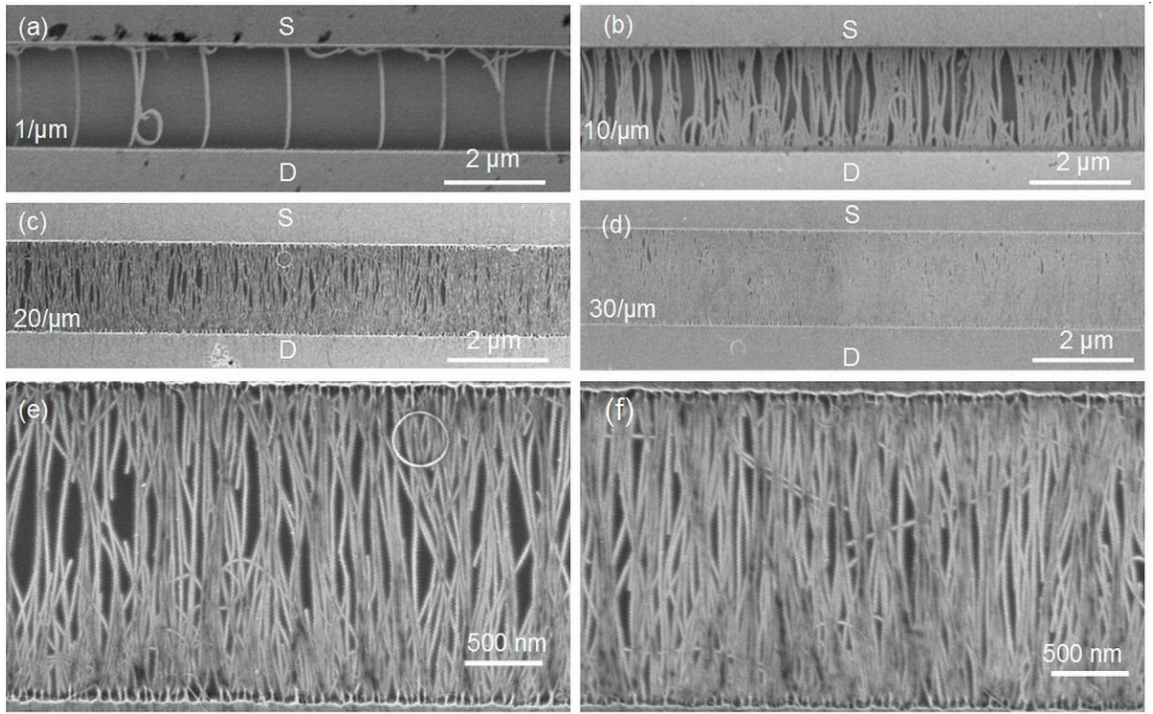

Figure 2. Scanning electron microscopy (SEM) image for a typical DEP assembly with varying nanotube density: (a) ~ 1 SWNT/μm. (b) 10 SWNT/μm. (c) 20 SWNT/μm and (d) 30 SWNT/ μm. (e) and (f) show a magnified image of (c) and (d). The nanotube density is varied by simply tuning the concentration of the solution while keeping all other DEP parameters fixed. For this assembly L=2 μm and W=25 μm.

We have also investigated the effect of further increase of SWNT concentration in the solution for this channel length. When the concentration of SWNT was increased to 6.8 μg/ml (8 times dilution) the linear density of the SWNT increased to more than 30 /μm as shown in figure 3a and 3b, however, we found that significant numbers of nanotubes form a second or even third layer making the nanotube film 2 to 3 layers thick at certain places as verified from AFM study. At 30 SWNT/μm, the inter nanotube separation is about 33 nm and it is not clear why the inter nanotube separation cannot be further decreased without making multi-layers. Further theoretical and experimental studies are required to elucidate this.

In order to demonstrate the scalability of the DEP assembly process, we have carried out similar alignment in larger channel length devices with L=5 μm and L=10 μm by varying the concentration of SWNT solution. This is shown in figure 4 where the left panel is for L= 5 µm and right panel is for L= 10 µm. The value of W was kept fixed at 25 µm. Figure 4a shows SEM image for the assembly when the concentration was 0.167 µg/ml (300 times dilution), which resulted in assembly of 20 SWNT on the entire 25 µm channel width giving a linear density of 0.8 SWNT/µm. Figure 4b and c show the alignment with density 8 and 28 SWNT/µm respectively. These alignments were achieved by the solution of concentration 0.83 µg/ml (60 times dilution) and 8.3 µg/ml (6 times dilution). The intermediate density alignments were also achieved by DEP using intermediate solution concentrations (images are not shown here). A small number of short nanotubes can also be seen at the bases of electrodes. This is due to the fact that assembled nanotubes into the circuit modify the electric field lines near to the electrode nanotube junctions and allow smaller nanotubes to gather around them [36]. In order to avoid such short nanotubes at the base, it will be necessary to find techniques where the smaller nanotubes can be filtered out from the solution. To further investigate the nature of alignment in large electrode separation, assembly was carried out on the device of L=10 µm. Figure 4d shows the SEM image of alignment with density 0.5 SWNT/µm. The concentration used for this assembly was 0.34 µg/ml (150 times dilution). A high density alignment (~ 18 SWNT/µm) is shown in the figure 4e and f. To achieve this alignment original solution is diluted by 10 times (5 µg/ml). The quality and density of alignment of SWNT do not exhibit any noticeable differences as L was increased from 2 µm to 10 µm. From these images, it can be seen that the electrode gap is mostly bridged by individual nanotubes irrespective of the different electrode gaps. It seems like that the DEP process favors the alignment of the nanotube of length comparable to L, although in larger channel length devices; a few of them are bridged by multiply connected nanotubes.

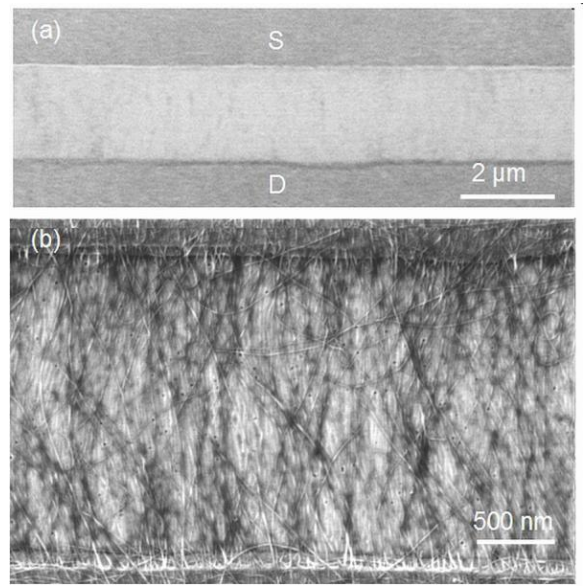

Figure 3. (a) SEM image of a densely packed array. (b) Magnified image to show that some nanotubes form $2^{nd}$ or $3^{rd}$ layer.

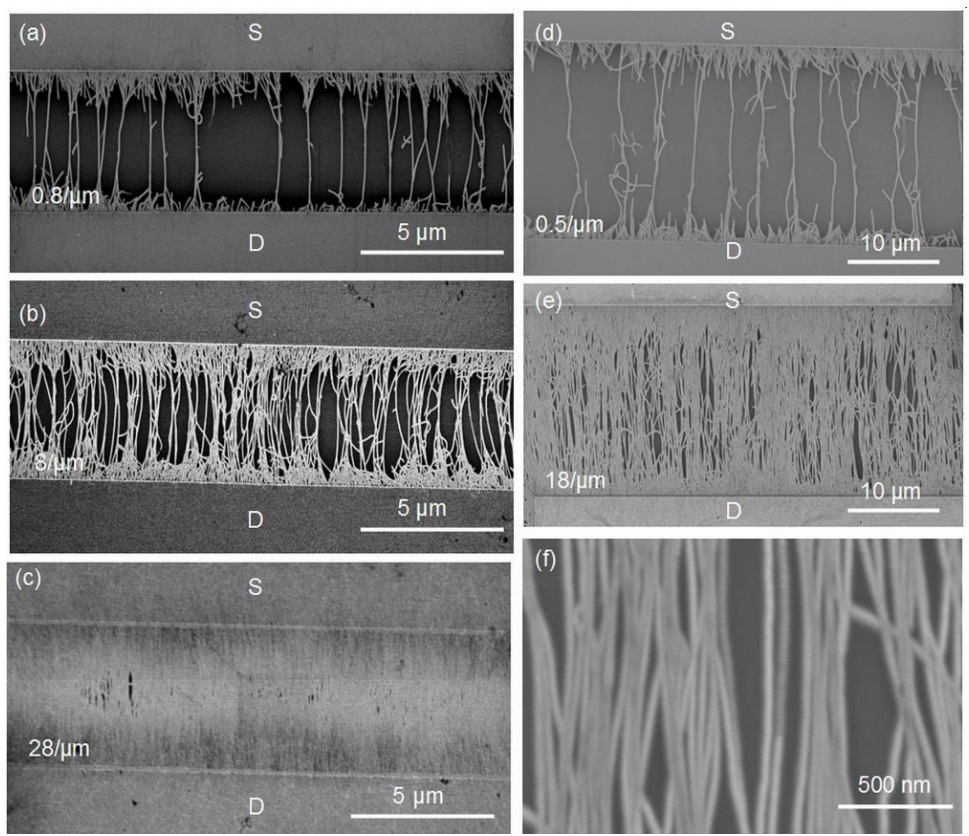

Figure 4. DEP assembly using higher channel length of 5 and 10 µm with tunable nanotube density. (a)-(c) SEM image showing ~ 1 SWNT/µm, 8 SWNT/µm and 28 SWNT/µm in 5 µm channel length device. (d)-(e) SEM image showing ~ 0.5 SWNT/µm and 18 SWNT/µm. (f) magnified image of (e).

Our DEP assembly technique can also be extended to homogeneously align SWNT for large channel width devices. Figure 5 shows the aligned array of nanotubes with SWNT density of 20 SWNT/µm on various channel width devices of W= 100 , 500 and 1000 µm. We found that the linear density of SWNT is independent of W when all DEP parameters were kept fixed except that the volume of the solution droplet needs to be increased. We noticed that the assembly is fairly uniform over the entire channel width.

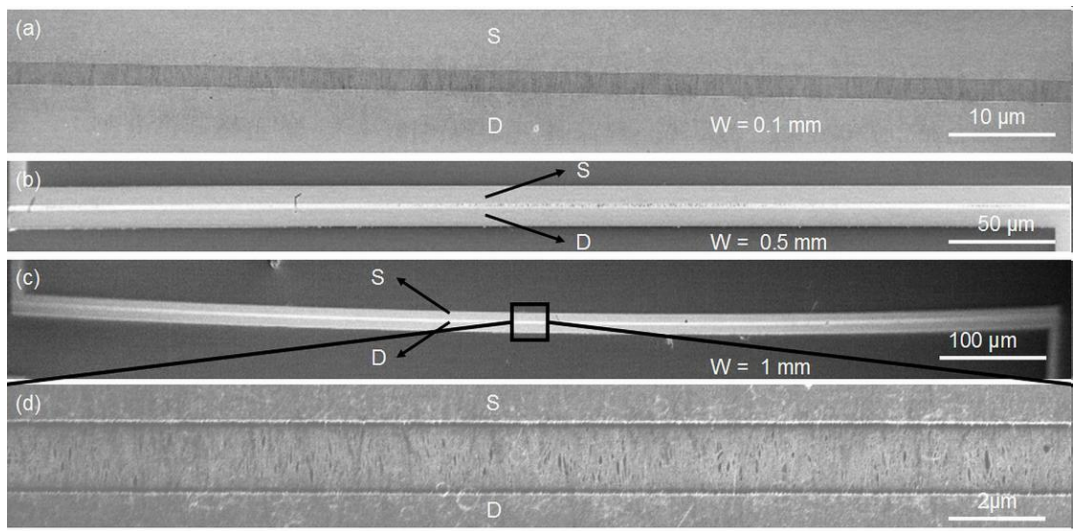

Figure 5. SEM image showing homogeneous assembly of SWNT on large channel width devices of L= 2 µm (a) W = 0.1 mm (100 µm). (b) W= 0.5 mm (500 µm) (c) W= 1 mm (d) Enlarge view of (c). Average density of CNT is 20/µm in all the images.

Figure 6 summarizes the density of nanotubes in aligned arrays vs. concentration of the SWNT solution for different channel lengths of L= 2, 5 and 10 µm. The DEP parameters were fixed but the effective electric field reduces with increasing L. For this reason, to achieve comparable alignment in L=5, and L=10 µm with that of L=2 µm device, a higher concentration of SWNT is required. For example, to achieve the density of 4 SWNT/µm on L=2 µm, L=5 µm and L=10 µm, the concentration of SWNT solution requirement varies as 0.167, 0.34 and 0.86 µg/ml respectively. There seems to be a tendency that with increasing concentration the linear density of SWNT in the array initially increases rapidly and become much slower at higher densities (saturation in log-log scale). After about 30 SWNT/µm, we observe formation of double layers. It is interesting to note that even in double growth CVD method; the maximum density did not exceed 30 SWNT/µm [18].

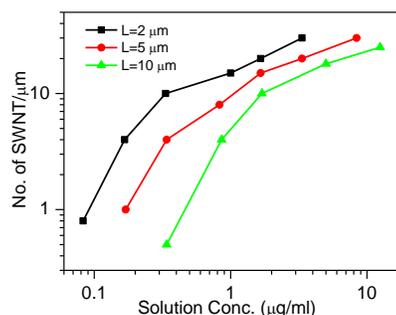

Figure 6. Variation of the linear density of SWNT in the array during DEP assembly with concentration of SWNT solution for different channel length devices.

We have examined the electrical characteristics of the assembled devices. Before electrical characterization, the devices were annealed at 200 $^0$C in Ar/H$_2$ atmosphere for an hour. Figure 7a shows the typical current-voltage (*I-V*) characteristics for a high density (20 SWNT/µm) alignment. At low bias the *I-V* curves are linear for all channel length and make ohmic contact with Pd electrodes. The resistance (R) values were found to be 500 Ω, 1 KΩ and 1.5 KΩ respectively for L=2 µm, L=5 µm and L =10 µm respectively. Figure 7b shows a plot of device resistance (after annealing) versus channel length for low, medium and high packing density of SWNT. For the same packing density, the R increases almost linearly with increasing channel length. R per unit length remains constant for all the cases. Since, the device resistance is the sum of contact resistance (R$_{contact}$) and channel resistance (R$_{channel}$), we can estimate both parameters from these curves. The contact and sheet resistance is related by the equation $R = R_{contact} + R_{channel}$ [11], where R$_{channel}$ = R$_{sheet}$(L/W). We obtain values of R$_{contact}$ ~ 2315, 380, and 330 Ω for low (1/µm), medium (4/µm), and high densities (20/µm) respectively. The sheet resistance is calculated from the slope of linear fits and their values were 8.8, 8.1, and 2.9 KΩ□$^{-1}$ for low, medium, and high densities respectively. The low sheet resistance demonstrates that our aligned array has potential application as electrode materials for different applications such as organic electronics and photovoltaic [12, 29].

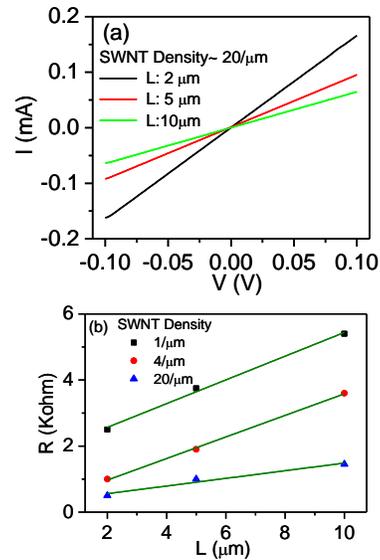

Figure 7. (a) I-V curves for SWNT array with ~ 20 SWNT/µm. (b) The resistance of the array versus channel length L for different SWNT density.

In previous DEP studies of SWNT, it was claimed that DEP assembly favors metallic (m) SWNT over semiconducting (s) SWNT. The low sheet resistance therefore raises question whether, the assembly contains any s-SWNT. In order to investigate this, we have used selective removal of metallic nanotubes using electrical breakdown technique and checked if field effect transistors (FET) can be fabricated using the remaining s-SWNTs. Figure 8a shows the representative *I-V* plot for three sequential breakdowns for one of our devices with 20 SWNT/µm and L= 5 µm. The back gate was held constant at $V_G$= +30V to deplete the carriers in the p-type semiconducting SWNTs while we ramped up V to eliminate the metallic SWNTs. As *V* was ramped up, the SWNTs started to breakdown and *I* began to fall. In order to obtain reproducible results, each breakdown was stopped when *I* had fallen by an order of magnitude and at that point V was swept back to zero. Figure 8b shows the variation of *I* as a function of $V_G$ after each breakdown. The bias (*V*) was kept constant at 0.5 V. The mobility (µ) was calculated by using the formula µ= (L/WVC) x (dI/dV$_G$) where C is the capacitance per unit area between the array and the gate. The as assembled array shows very little gate modulation with current on-off ratio of 1.5 and corresponding mobility of ~50 cm$^2$/Vs due to the presence of large number of metallic pathways. After the first breakdown the FET behavior of the device is enhanced with on-off ratio ~ 10 and µ~ 30 cm$^2$/Vs. After the 2$^{nd}$ breakdown the on-off ratio increased to ~100 with µ~20 cm$^2$/Vs. Finally, after the 3$^{rd}$ breakdown the on-off ratio increases to ~ 1.3×10$^4$ with µ ~10 cm$^2$/Vs.

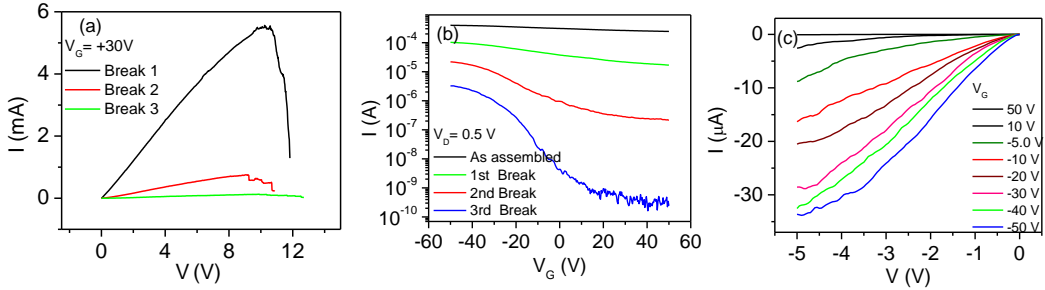

Figure 8. (a) I - V characteristics for three sequential electrical breakdowns. I was allowed to fall by an order of magnitude during the first two breakdown after which the voltage was swept back to zero. (b) I vs.$V_G$ (back gate voltage) at constant V of 0.5 V after each breakdown. The on-off ratio improves after each breakdown due to removal of metallic pathways. (c) Output characteristics of the FET after third breakdown.

Figure 8c shows the output characteristics, $I$ versus $V$ at different $V_G$ recorded for the same sample after the third breakdown. A more systematic study of FET properties with different density in the array is under investigation. However, this result clearly shows that both m-SWNT and s-SWNT is present in our aligned array devices and that m-SWNT can be selectively removed to fabricate FET.

In conclusion, we have demonstrated ultra high density alignment of SWNT in an array using high quality SWNT aqueous solution and dielectrophoresis. By tuning the concentration of the SWNT solution, we can control the density of the nanotube up to 30 SWNT/μm in the array which is the highest reported density using any solution processing technique. Effort to further increase the density resulted multiple layers of nanotubes in the array. Electrical measurement data shows that the densely packed aligned arrays have low sheet resistances. Selective removal of metallic SWNTs via controlled electrical breakdown produced field effect transistors (FET) with high current on-off ratio. Ultra-high density alignment using solution processed SWNT reported here will have important implication in fabricating high quality nanoelectronic devices.

**MATERIAL AND METHODS:** Devices were fabricated on heavily doped silicon substrates capped with a thermally grown 250 nm thick $SiO_2$ layer. The electrode patterns were fabricated by a combination of optical and electron beam lithography (EBL). First, contact pads and electron beam markers were fabricated with optical lithography using double layer resists (LOR 3A/Shipley 1813) developing in CD26, thermal evaporation of 3 nm chromium (Cr) and 50 nm gold (Au) followed by lift-off. Smaller electrode patterns were fabricated with EBL using single layer PMMA resists and then developing in (1:3) methyl isobutyl ketone:isopropal alchohol (MIBK:IPA). After defining the patterns, 3 nm Cr and 27 nm thick Pd were deposited using electron beam deposition followed by lift-off in boiling acetone. Pd was used because it is known to make the best electrical contact to SWNTs. The channel lengths (L) were 2, 5, and 10 μm while the channel width (W) was varied from 25 μm to 1 mm. There are 9 independent drain and source (electrodes) pairs on each chip.

The devices were imaged using a Zeiss Ultra 55 field emission scanning electron microscope. The microscope is capable of delivering high lateral resolution at low voltages. Inlens detector was used to get image at low voltages. The electrical measurements were performed in a probe station by a high resolution DAC card and a current pre-amplifier (DL1211) interfaced with LabView.